# Weak 'antigravity' fields in general relativity

F. S. Felber[a)]

*Physics Division, Starmark, Inc., P. O. Box 270710, San Diego, California 92198*

Within the weak-field approximation of general relativity, new exact solutions are derived for the gravitational field of a mass moving with arbitrary velocity and acceleration. A mass having a constant velocity greater than $3^{-1/2}$ times the speed of light gravitationally repels other masses at rest within a narrow cone. At high Lorentz factors ($\gamma \gg 1$), the force of repulsion in the forward direction is about $-8\gamma^5$ times the Newtonian force, offering opportunities for laboratory tests of gravity at extreme velocities. One such experiment is outlined for the Large Hadron Collider.

PACS Numbers: 04.20.Jb, 04.25.–g, 04.20.–q

### I. Introduction

This paper calculates in the weak-field approximation of general relativity the retarded gravitational field in conventional 3-vector notation of a particle mass having any arbitrary motion. Earlier calculations [1–5] of the gravitational fields of arbitrarily moving particles were done only to first order in $\boldsymbol{\beta} = \mathbf{u}/c$, the ratio of source velocity to the speed of light. Because the affine connection is *non-positive-definite*, a "general prediction" had been made that general relativity could admit a repulsive force at relativistic speeds [6]. But since the repulsive-force terms are second-order and higher in $\boldsymbol{\beta}$, this 'antigravity' had not previously been found by this or any other approach.

We recently derived and analyzed *exact* time-dependent field solutions of Einstein's gravitational field equation for a spherical mass moving with arbitrarily high *constant* velocity [7]. For the special case of a mass moving with constant velocity, the exact 'antigravity' fields calculated in [7] correspond at all source speeds to the weak 'antigravity' fields for any arbitrary velocity and acceleration derived and analyzed by a retarded-potential methodology in this paper and in [8,9]. This paper presents the explicit coordinate transformations that demonstrate this correspondence.

The exact dynamic fields in [7] were calculated from an exact metric, which was first derived by Hartle, Thorne, and Price [10], but which had never before been analyzed. The exact 'antigravity-field' solutions in [7] confirm that any mass having a speed greater than $3^{-1/2}c$ gravitationally repels particles at rest within a forward and backward cone, no matter how light the mass or how weak its field.

The relativistically exact bound and unbound orbits of test particles in the strong *static* field of a mass at rest have been thoroughly characterized in [1,4,11], for example. Even in a weak static field, earlier calculations of fields [1–5] only solved the geodesic equation for a *nonrelativistic* test particle in the *slow-velocity* limit of source motion. In this slow-velocity limit, the field at a *moving* test particle has terms that look like the Lorentz field of electromagnetism, called the 'gravimagnetic' or 'gravitomagnetic' field [3–5]. Harris [3] derived the *nonrelativistic* equations of motion of a moving test particle in a dynamic field, but only the dynamic field of a *slow-velocity* source. Mashhoon [12] calculated the dynamic gravitomagnetic field of a slowly spinning source having slowly varying angular momentum, and more generally showed explicitly that general relativity contains induction effects at slow source velocities [13].

An exact solution of the field of a relativistic mass is the Kerr solution [1,4,11,14], which is the exact *stationary* solution for a rotationally symmetric spinning mass. Although time-independent, the Kerr field exhibits an inertial-frame-dragging effect [1] similar to that contributing to gravitational repulsion at relativistic velocities. In the stationary Kerr gravitational field, the relativistic unbound orbits of test particles have been approximated in [15].

This paper calculates and analyzes the first relativistically exact nonstationary field within the weak-field approximation of general relativity. A Liénard-Wiechert "retarded solution" approach [16] is used to solve the linearized field equations in the weak-field approximation from the retarded Liénard-Wiechert tensor potential of a relativistic particle. The solution is used to calculate the weak field acting on a test particle at rest of a mass moving with arbitrary relativistic motion. Since this solution is the first ever used to analyze the field of a translating mass beyond first order in $\boldsymbol{\beta}$, it is the first to reveal that a mass having a constant velocity greater than $3^{-1/2}c$ gravitationally repels other masses at rest within a cone, as seen by a distant inertial observer. (The Aichelburg-Sexl solution [17] and other boosted solutions, such as in [15], apply only in the proper frame of the accelerating particle, in which no repulsion appears, and not in the laboratory frame.)

That a particle with a radial speed exceeding $3^{-1/2}c$ is repelled in a weak static Schwarzschild field may first have been correctly noted by Hilbert [18] in 1924. Subsequent papers have addressed [19] and reviewed [20] the critical speed for radial motion in a weak Schwarzschild field. The same critical speed of repulsion was found for radial motion along the rotation axis of a spinning stationary source in [21].

The discovery in this paper of a repulsive weak gravitational field produced by relativistic masses was used in [8] to calculate the exact relativistic motion of a particle in the strong gravitational field of a mass moving with constant relativistic velocity, but without an explicit calculation of the strong dynamic gravitational field that produced the motion. References [8] and [9] then showed how even a weak repulsive field of a suitable driver mass at relativistic speeds could quickly propel a heavy payload from rest to a speed significantly faster than the driver and close to the speed of light, and do so with manageable stresses on the payload.

### II. Weak 'Antigravity' Fields

Kopeikin and Schäfer [16] recognized that solving for the *dynamic* field of a relativistic particle in arbitrary motion requires a Liénard-Wiechert "retarded solution" approach. By using the retarded Green's function to solve the linearized field equations in the weak-field approximation, they calculated the exact "retarded Liénard-Wiechert tensor potential" of a relativistic particle of rest mass $m$ [16],

---

[a)]Electronic mail: felber@san.rr.com

$$h_{\mu\nu}(\mathbf{r},t) = \frac{-4Gm}{c^4} \int \frac{S_{\mu\nu}(t')}{\gamma(t')R(t')} \delta\left(t' + \frac{R(t')}{c} - t\right) dt'$$
$$= \frac{-4Gm}{c^4} \left\{ \frac{S_{\mu\nu}}{\gamma \kappa R} \right\}_{ret} . \quad (1)$$

In the weak-field approximation used in Eq. (1), the metric tensor was linearized as $g_{\mu\nu} = \eta_{\mu\nu} + h_{\mu\nu}$, where $\eta_{\mu\nu}$ is the Lorentz metric; $S_{\mu\nu} \equiv u_\mu u_\nu - c^2 \eta_{\mu\nu}/2$ is a source tensor, with pressure and internal energy neglected; $u_\mu = \gamma(c, \mathbf{u})$ is the 4-velocity of the source; $\mathbf{u}$ is the 3-velocity; $\gamma = (1-\beta^2)^{-1/2}$ is the Lorentz (relativistic) factor; $\mathbf{R} = \mathbf{r} - \mathbf{r}'$ is the displacement vector from the source position $\mathbf{r}'(t')$ to the observation point $\mathbf{r}(t)$; $\mathbf{n} = \mathbf{R}/R$ is a unit vector; the delta function provides the retarded behavior required by causality; the factor $\kappa \equiv 1 - \mathbf{n} \cdot \mathbf{u}/c$ is the derivative with respect to $t'$ of the argument of the delta function, $t' + [R(t')/c] - t$; and the quantity in brackets $\{\ \}_{ret}$ and all primed quantities are to be evaluated at the retarded time $t' = t - R'/c$.

Equation (1) is the starting point in this paper for the exact calculation of the gravitational field of a relativistic particle from the tensor potential. In the weak-field approximation, the retarded tensor potential of Eq. (1) is exact, even for relativistic velocities of the source. And since the tensor potential is linear, the field to be derived from it in this paper is easily generalized to ensembles of particles and to continuous source distributions. A treatment of post-linear corrections to the Liénard-Wiechert potentials in [16] is given in [22] and references therein.

To derive an exact expression for the gravitational field from Eq. (1) by the Liénard-Wiechert formalism [23], we define a 'scalar potential,'

$$\Phi(\mathbf{r},t) \equiv \frac{c^2 h_{00}}{2} = -Gm \int \frac{\alpha'}{R'} \delta\left(t' + \frac{R'}{c} - t\right) dt' , \quad (2)$$

and a 'vector potential' having components,

$$A^i(\mathbf{r},t) \equiv c^2 h_0^i = -4Gm \int \frac{(\beta^i)' \gamma'}{R'} \delta\left(t' + \frac{R'}{c} - t\right) dt' , \quad (3)$$

where $\alpha \equiv 2\gamma - 1/\gamma$, and $i = 1,2,3$. Then from the geodesic equation, the equation of motion of a test particle *instantaneously at rest* at $(\mathbf{r},t)$ in a weak field is

$$\frac{d^2\mathbf{r}}{dt^2} = -\nabla \Phi(\mathbf{r},t) - \frac{1}{c} \frac{\partial \mathbf{A}(\mathbf{r},t)}{\partial t} . \quad (4)$$

Since the gradient operation in Eq. (4) is equivalent to $\nabla \to \mathbf{n} \partial/\partial R$, the contribution of the 'scalar potential' to the gravitational field can be written as

$$-\nabla \Phi = -Gm \int \left[ \frac{\alpha \mathbf{n}}{R^2} \delta\left(t' + \frac{R'}{c} - t\right) - \frac{\alpha \mathbf{n}}{cR} \dot\delta\left(t' + \frac{R'}{c} - t\right) \right] dt' . \quad (5)$$

The contribution of the 'vector potential' is

$$-\frac{1}{c} \frac{\partial \mathbf{A}}{\partial t} = \frac{-4Gm}{c} \int \frac{\gamma \boldsymbol{\beta}}{R} \dot\delta\left(t' + \frac{R'}{c} - t\right) dt' , \quad (6)$$

where $\dot\delta$ is a delta function differentiated with respect to its argument. If the variable of integration is changed to $f(t') = t' + R'/c$, then integrating by parts on the derivative of the delta function, and combining Eqs. (4) to (6) gives

$$\mathbf{g}(\mathbf{r},t) = -Gm \left\{ \frac{\alpha \mathbf{n}}{\kappa R^2} + \frac{1}{c\kappa} \frac{d}{dt'} \left( \frac{\alpha \mathbf{n} - 4\gamma \boldsymbol{\beta}}{\kappa R} \right) \right\}_{ret} . \quad (7)$$

To calculate the time derivatives in Eq. (7), the following relations from [10] are used,

$$\frac{1}{c} \frac{d\mathbf{n}}{dt'} = \frac{\mathbf{n} \times (\mathbf{n} \times \boldsymbol{\beta})}{R} = \frac{(\mathbf{n} \cdot \boldsymbol{\beta})\mathbf{n} - \boldsymbol{\beta}}{R} , \quad (8)$$

$$\frac{1}{c} \frac{d(\kappa R)}{dt'} = \beta^2 - \mathbf{n} \cdot \boldsymbol{\beta} - \frac{R(\mathbf{n} \cdot \dot{\boldsymbol{\beta}})}{c} , \quad (9)$$

where an overdot denotes differentiation with respect to $t'$. Applying these relations, Eqs. (8) and (9), to Eq. (7) gives the relativistically exact (weak) retarded gravitational field of a moving source on a test particle instantaneously at rest at $(\mathbf{r},t)$ as

$$\mathbf{g}(\mathbf{r},t) = -Gm \left\{ \frac{(\alpha/\gamma^2)\mathbf{n} + [(2\gamma + 1/\gamma)\kappa - 4/\gamma]\boldsymbol{\beta}}{\kappa^3 R^2} \right.$$
$$\left. + \frac{(\mathbf{n} \cdot \dot{\boldsymbol{\beta}})(\alpha \mathbf{n} - 4\gamma \boldsymbol{\beta}) + \kappa(\dot\alpha \mathbf{n} - 4\dot\gamma \boldsymbol{\beta} - 4\gamma \dot{\boldsymbol{\beta}})}{c\kappa^3 R} \right\}_{ret} . \quad (10)$$

Just as does the retarded electric field of a point charge [10], the retarded gravitational field divides itself naturally into a 'velocity field,' which is independent of acceleration and which varies as $R^{-2}$, and an 'acceleration field,' which depends linearly on $\dot\beta$ and which varies as $R^{-1}$. The radial component of the electric 'velocity field' never changes sign. The radial component of the gravitational 'velocity field,' on the other hand, can change sign at sufficiently high source velocity and repel masses within a narrow cone.

Equation (10) is the gravitational field that would be observed by a distant inertial observer to act on a test particle at rest at $(\mathbf{r},t)$. If the test particle moves with velocity $\mathbf{v}$, then the gravitational field measured by the moving test particle has additional 'gravimagnetic' terms. The following calculations apply in the rest frame of the test particle and use the impulse approximation, which has the test particle remaining at rest during the time the gravitational field of the moving particle acts upon it. (Any difference in fields caused by motion induced in the test particle by the source is of the same order as terms that have already been neglected in the weak-field approximation.)

A particle of mass $m_0$ moving with constant velocity $\boldsymbol{\beta}_0 c$ and $\gamma_0 = (1-\beta_0^2)^{-1/2}$ at impact parameter $b$ with respect to a test particle at rest at $(\mathbf{r},t)$, as shown in Fig. 1, produces a gravitational field at the test particle with a radial component,

$$g_R = \mathbf{g} \cdot \mathbf{n} =$$
$$-Gm_0 \left\{ \frac{1 - \beta_0^4 - [1 - 3\beta_0^2 + (3 - \beta_0^2)\beta_0 \cos\theta]\beta_0 \cos\theta}{(1-\beta_0^2)^{1/2}(1-\beta_0 \cos\theta)^3 R^2} \right\}_{ret} . \quad (11)$$

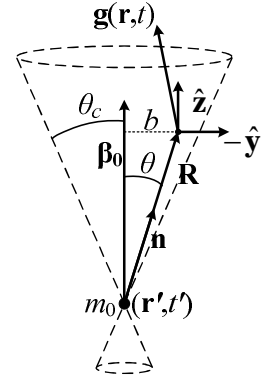

FIG. 1. Configuration of mass $m_0$ moving with constant $\boldsymbol{\beta}_0$ and repelling a test particle at rest at $(\mathbf{r},t)$, impact parameter $b$, and polar angle $\theta$ within critical half-angle $\theta_c$ of forward cone.

This radial component is plotted in Fig. 2 for several values of the retarded polar angle $\theta'$ defined in Fig. 1. Figure 2 shows that for sufficiently narrow polar angles $\theta'$, and for sufficiently high $\beta_0$, $g_R$ reverses direction and repels stationary particles.

Figure 3 shows as a function of $\beta_0$ the retarded critical polar angle,

$$\theta_c' = \cos^{-1}\left[\frac{3\beta_0^2 - 1 + (13 - 10\beta_0^2 - 3\beta_0^4 + 4\beta_0^6)^{1/2}}{2\beta_0(3 - \beta_0^2)}\right], \quad (12)$$

inside which $g_R$ repels stationary masses, instead of attracting them. Analysis of Eq. (12) shows that $\beta_0 > 3^{-1/2}$ and either $\theta' < 23.844$ deg or $\theta' > 180 - 23.844$ deg are both necessary conditions for gravitational repulsion of stationary masses by a particle having constant velocity. For $\gamma_0 \gg 1$, the 'antigravity beam' divergence narrows to $\theta_c' \approx 1/\gamma_0$.

If a mass $m$ approaching or receding from a stationary test particle at a speed greater than $3^{-1/2}c$ repels the test particle, then a stationary mass $M$ should also repel a test particle that approaches or recedes from it at a speed greater than $3^{-1/2}c$. In the Schwarzschild field of a stationary mass $M$, the exact equation of motion of a test particle having purely radial motion (zero impact parameter) is

$$\frac{d^2R}{dt^2} = -\frac{GM}{R^2}\left[\psi - \frac{3}{\psi c^2}\left(\frac{dR}{dt}\right)^2\right], \quad (13)$$

where $\psi(R) \equiv g_{00} = 1 - 2GM/Rc^2$. In the weak-field approximation, $\psi \approx 1$, and a test particle approaching or receding from the stationary mass $M$ at a speed greater than $3^{-1/2}c$ is repelled, a result first derived by Hilbert [18].

For $\gamma \gg 1$, Eq. (11) shows that $g_R \approx -8\gamma^5 g_N$ in the forward direction ($\theta' = 0$), where $g_N = -Gm_0/\{R^2\}_{\text{ret}}$ is the Newtonian field. For $\gamma \gg 1$, the repulsive field in the backward direction ($\theta' = \pi$), $g_R \approx -(\gamma/2)g_N$, is much weaker. At extreme relativistic velocities, such as are attained in particle accelerators and storage rings, the repulsive gravitational field of particle bunches in their forward direction can be many orders of magnitude greater than the Newtonian field.

### III. Exact 'Antigravity' Fields

The exact dynamic metric of a mass $m$, spherically symmetric in its rest frame, moving with constant velocity $\beta_0 c$ in the $+z$ direction was first derived, but not analyzed, in [10]. The only nonvanishing components of the metric in $t$, $x$, $y$, $z$ coordinates are

$$g_{00} = p - \beta_0^2 q, \quad g_{03} = g_{30} = -\beta_0(p - q),$$
$$g_{11} = g_{22} = -q/\gamma_0^2, \quad g_{33} = \beta_0^2 p - q, \quad (14)$$

where $p \equiv \gamma_0^2[(1-\rho)/(1+\rho)]^2$, $q \equiv \gamma_0^2(1+\rho)^4$, $\rho \equiv r_0/\tilde{r}$, $r_0 = Gm/2c^2$, and $\tilde{\mathbf{r}} \equiv \tilde{\mathbf{x}} + \tilde{\mathbf{y}} + \tilde{\mathbf{z}}$ is the displacement vector from the source position to the observation point $\tilde{\mathbf{r}}(\tilde{t})$ in the 'comoving' Cartesian coordinates,

$$\tilde{t} = \gamma_0(t - \beta_0 z/c), \quad \tilde{x} = x, \quad \tilde{y} = y, \quad \tilde{z} = \gamma_0(z - \beta_0 ct). \quad (15)$$

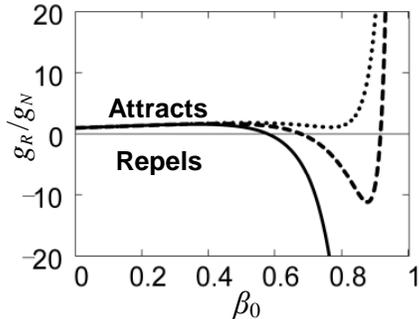

FIG. 2. Radial component of gravitational field normalized to Newtonian field, $g_R/g_N$, vs. constant $\beta_0$ of source at retarded polar angles $\theta'$ of 0 deg (solid), 20 deg (dashed), 25 deg (dotted).

A simple calculation of an exact strong dynamic gravitational field from Eq. (14) is the field on a sensor at rest from a mass $m$ moving along the $z$ axis. Since the 3-velocity of the sensor is zero, the exact equations of motion from the geodesic equation are $dt/d\tau = (g_{00})^{-1/2}$ where $\tau$ is the proper time, and, in coordinate time $t$, $d^2x^i/dt^2 + c^2\Gamma^i_{00} = 0$, where $i = 1, 2, 3$, and the $\Gamma^i_{00}$ are Christoffel symbols. The exact strong dynamic gravitational field on a sensor at rest at $(t, x, y, z)$, as seen by a *distant inertial observer*, is therefore

$$\mathbf{g}(\mathbf{r},t) = \frac{d^2\mathbf{r}}{dt^2} = -\frac{\gamma_0^2 Gm}{\tilde{r}^3}\left\{\left(\frac{1-\rho}{(1+\rho)^7} + \frac{\beta_0^2}{1+\rho}\right)(\tilde{\mathbf{x}} + \tilde{\mathbf{y}}) + \left[\frac{1-\rho}{(1+\rho)^7} - \beta_0^2\left(\frac{3-\rho}{1-\rho^2}\right)\right]\gamma_0\tilde{\mathbf{z}}\right\}. \quad (16)$$

This exact gravitational field is derivable from scalar and vector potentials,

$$\phi(\rho) = -2\gamma_0^2 c^2\left[\frac{2}{15} - \frac{2 - 3\rho}{15(1+\rho)^6} + \beta_0^2 \ln(1+\rho)\right], \quad (17)$$
$$\mathbf{a}(\rho) = -2\gamma_0^2 c^2 \ln\left[(1+\rho)^3/(1-\rho)\right]\boldsymbol{\beta_0}$$

which satisfy the Lorentz gauge condition [23] for the 3-vector field equation $\mathbf{g} = -\nabla\phi - (1/c)\partial\mathbf{a}/\partial t$. In the weak-field approximation ($\rho \ll 1$), the potentials are $\phi \approx -(1+\beta_0^2)\gamma_0^2 Gm/\tilde{r}$ and $\mathbf{a} \approx -4\gamma_0^2 Gm\boldsymbol{\beta_0}/\tilde{r}$.

Equation (16) is an exact solution of Einstein's equation for the field on a particle at rest produced by a mass moving along the $z$ axis. In the weak-field approximation ($\rho \approx 0$), Eq. (16) agrees with Eq. (10), and the 'antigravity' threshold on the $z$ axis is $\beta_0 \approx 3^{-1/2}$. The velocity threshold for 'antigravity' is reduced in strong fields [7]. Equation (16) also confirms the weak-field result that the transverse field component is always attractive, meaning that a particle at rest will only be repelled by a relativistic mass if the particle lies within a sufficiently narrow angle to its forward or backward direction. That is, to be repelled, the particle must lie within the conical surface on which $g$ vanishes. In the weak-field approximation, this critical half-cone angle for the 'antigravity' threshold, measured from the $z$ axis is

$$\theta_c = \tan^{-1}[(3\beta_0^2 - 1)/(1-\beta_0^4)]^{1/2} . \quad (18)$$

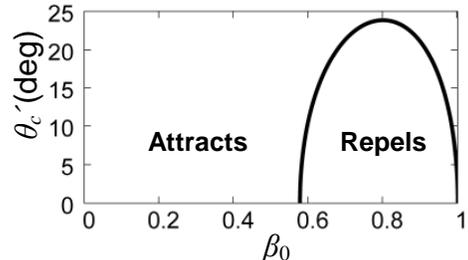

FIG. 3. Retarded critical half-angle of forward cone, within which radial field $g_R$ of source repels stationary masses, vs. constant $\beta_0$ of source.

This 'antigravity beam' angle in Eq. (18) is seen to be equivalent to the critical angle calculated in retarded coordinates in Eq. (12) by means of the transformations given in Table 1. Like the retarded critical angle in Eq. (12), in these coordinates the 'antigravity beam' narrows in the forward and backward directions for $\gamma_0 \gg 1$ to $\theta_c \approx 1/\gamma_0$.

| | |
|---|---|
| $t = t' + r'/c,$ | $t' = \gamma_0^2(t - \beta_0 z/c) - \gamma_0 \tilde{r}/c$ |
| $x = x', \quad y = y'$ | $x' = x, \quad y' = y$ |
| $z = z' + \beta_0 ct'$ | $z' = \gamma_0^2(z - \beta_0 ct) + \beta_0 \gamma_0 \tilde{r}$ |
| $z - \beta_0 ct = z' - \beta_0 r'$ | $z' + \beta_0 ct' = z$ |
| $\tilde{r} = \gamma_0(r' - \beta_0 z') = \gamma_0 \kappa' r'$ | $r' = \beta_0 \gamma_0^2(z - \beta_0 ct) + \gamma_0 \tilde{r}$ |
| $dt = \kappa' dt' = -\kappa' dz'/\beta_0 c$ | $dt'/r' = \gamma_0 dt/\tilde{r}$ |
| $dz = 0$ | $dz'/r' = -\beta_0 \gamma_0 cdt/\tilde{r}$ |

Table 1. Transformations to and from retarded (primed) coordinates.

With these transformations, and in the weak-field approximation, Eq. (16) expressed in retarded coordinates becomes
$$\mathbf{g}(\mathbf{r},t) \approx -\left(\gamma_0 Gm/(\kappa' r')^3\right)\left[(1-\beta_0^4)(\mathbf{x}'+\mathbf{y}') + (1-3\beta_0^2)(\mathbf{z}' - r'\boldsymbol{\beta_0})\right] . \quad (19)$$
This field is identical to the 'velocity field' in Eq. (10) for the special case of constant source velocity along the $z$ axis. Correspondence is thereby demonstrated between the weak 'antigravity' field derived from retarded potentials in Eq. (10) and the exact 'antigravity' field in Eq. (16) derived from an exact metric in [7].

### IV. Impulse Approximations

As a simple application of the field derived in Eq. (16), we will now calculate to first order in the field-strength parameter, $\varepsilon \equiv GM/bc^2$, the angular deflection of a particle of mass $m$ moving along the $z$ axis at nearly constant speed $\beta_0 c$ in the weak static Schwarzschild field of a much larger mass $M$, located at $z=0$, $y=b$. In the weak-field impulse approximation ($\varepsilon \ll \beta_0^2$) of Eq. (16), the field on $M$ is
$$g_y \approx -(1+\beta_0^2)\gamma_0^2 Gmb\left[(\gamma_0 \beta_0 ct)^2 + b^2\right]^{-3/2} . \quad (20)$$
The transverse impulse delivered to the mass $M$ is
$$P_y \approx M\int_{-\infty}^{+\infty} g_y dt \approx -2(1+1/\beta_0^2)\varepsilon P_0 , \quad (21)$$
where $P_0 = \gamma_0 mc\beta_0$ is the momentum of the particle. Since $P_y$ is equal and opposite to the impulse delivered to the particle, the angular deflection of the particle is $2(1+1/\beta_0^2)\varepsilon$, a result derived in [9] by integrating the orbit equation in a Schwarzschild field. This deflection corresponds to the deflection of a photon in a weak field for $\beta_0 = 1$.

In the weak-field impulse approximation, the *maximum* longitudinal impulse delivered to the mass $M$ is
$$P_z \approx M\int_{-\infty}^{0} g_z dt \approx +(3-1/\beta_0^2)\gamma_0 \varepsilon P_0 . \quad (22)$$
An equal and opposite longitudinal impulse is delivered to the particle. The same transverse and longitudinal impulses can be calculated in retarded coordinates by integrating Eq. (19) over $dt = \kappa' dt'$.

### V. Test of Relativistic Gravity at LHC

The strong dependence of longitudinal impulse on $\gamma$ offers opportunities for laboratory tests of gravity at relativistic velocities. Several methods should be able to provide accurate impulse measurements for the purpose of testing general relativity and discriminating among competing theories of gravity. For example, as suggested in [2], periodic gravitational impulses delivered by proton bunches circulating in a storage ring could be measured by detectors resonant at the bunch frequency. Because the impulse scales for $\gamma \gg 1$ as $\gamma^4$, and because the Large Hadron Collider (LHC) is much more powerful than the Tevatron-scale collider considered in [2], the signal strengths estimated here are orders of magnitude higher, and the experiment is much more feasible now, than projected by [2] over 30 years ago.

Comparing the phase of a stress wave in the detector to the phase of a proton bunch in the ring could give the first direct evidence of gravitational repulsion. Such an experiment to detect 'antigravity' impulses for the first time and to test relativistic gravity can be performed at the LHC off-line at any point in the tunnel, causing no interference with normal operations. For maximum effect, the resonant detector should be positioned in the plane of the proton ring at a distance $b$ as close as practical to the beam axis, so that the 'antigravity beam' sweeps across the detector at the closest range.

In this configuration, the 'antigravity beam' irradiates the detector in the forward direction of the proton bunch when the bunch is a retarded distance of $r' \approx (2bR_0)^{1/2}$ from the detector, where $R_0$ is the ring radius. From Eq. (19), the peak repulsive field of the bunch on the detector at that distance is $g_z \approx 4\gamma_0^5 Gm/bR_0$. Since the divergence of the 'antigravity beam' is about $1/\gamma_0$, it sweeps across a point on the detector in a time $\Delta t \approx R_0/\gamma_0 c$, and the specific impulse delivered to the detector by a single bunch is $g_z \Delta t \approx 4\gamma_0^4 Gm/bc$. The 'antigravity beam' spot size at the detector is $r'/\gamma_0$.

For $n_b$ equally spaced proton bunches in the ring, the bunch frequency and the impulse frequency at the detector is $f = n_b f_0$, where $f_0 = c/2\pi R_0$ is the circulation frequency. The duty cycle of the impulse is about $f\Delta t$, and the effective root-mean-square (rms) 'antigravity' wave amplitude is $g_{rms} \approx g_z f\Delta t$. The effective rms pressure $P_{rms}$ at the base of the detector is $g_{rms}$ times the areal mass density. For a resonant detector $N$ wavelengths thick, $P_{rms} \approx NZg_z\Delta t$, where $Z$ is the characteristic acoustic impedance of the detector. The effective sound pressure level of the signal is $SPL = 20\log_{10}(P_{rms}/1 \text{ μPa})$ dB $re$ 1 μPa.

The change in velocity of the detector face produced by a single bunch is just the specific impulse, $g_z\Delta t$. The change in displacement amplitude of the face (of a resonant detector) produced by a single bunch is $z_0 \approx g_z\Delta t/2\pi f$. The steady-state resonant displacement amplitude is about $Qz_0$, where $Q$ is the quality factor of the detector oscillator. The steady-state resonant amplitude is neared only after about $Q$ resonant oscillations of the detector.

The nominal beam parameters of the LHC are $1.15\times 10^{11}$ protons per bunch at $\gamma_0 = 7461$, circulating at radius $R_0 = $ 4.243 km and frequency $f_0 = 11.245$ kHz [24]. For an impact parameter $b \approx 10$ cm, each bunch irradiates the detector with an 'antigravity beam' of amplitude $g_z \approx 3$ nm/s² at a standoff range of $r' \approx 30$ m with a spot size of about $r'/\gamma_0 \approx 4$ mm. (In comparison, [2] estimated a signal strength of order $10^{-21}$ m/s² for a Tevatron-scale collider.) The specific impulse delivered to a detector by a single bunch during exposure to the 'antigravity beam' for $\Delta t \approx 2$ ns is about $5 \times 10^{-18}$ m/s.

When all 2808 rf buckets are filled with proton bunches, the bunch frequency and impulse frequency at the detector is $f = 31.576$ MHz. The duty cycle is $f \Delta t \approx 0.06$. A quartz crystal has a characteristic acoustic impedance $Z = 15$ MPa·s/m. The effective $SPL$ of the 'antigravity beam' at the base of an $N$-wavelength-thick quartz detector, therefore, is $SPL \approx -80 + 20\log_{10}(N)$ dB $re$ 1 µPa.

Near steady-state resonance, the $SPL$ is amplified by about $20\log_{10}(Q)$ dB. Since in normal operation at LHC, the beam can circulate for 10 to 24 hours with a bunch frequency over 30 MHz, $Q$ could be as high as $10^{12}$. Such a $Q$ is well within the theoretical limits of sapphire monocrystals that [2] suggests could be used for such an experiment.

## VI. Conclusions

This paper derived, in the weak-field approximation of general relativity, the relativistically exact gravitational field of a particle having any velocity and acceleration. The exact solution is the first to show that a distant inertial observer sees particles with constant velocity greater than $3^{-1/2}c$ repel stationary masses within an 'antigravity beam' in the forward and backward directions. This result should perhaps not be surprising, since Hilbert [18] showed 85 years ago that a distant inertial observer sees a stationary mass repel other masses moving radially towards it or away from it with a velocity greater than $3^{-1/2}c$.

At high Lorentz factors ($\gamma \gg 1$), the force of repulsion in the forward direction is about $-8\gamma^5$ times the Newtonian force. The strong scaling of impulse with $\gamma$ should make certain laboratory tests of relativistic gravity much less difficult than earlier imagined [2]. This paper outlined one such test that could be performed off-line at the LHC without interfering with any of the operations of the facility. The experiment would measure the repulsive gravitational impulses of proton bunches delivered in their forward direction to resonant detectors just outside the beam pipe. This test could provide accurate measurements of post-Newtonian parameters and the first observation of 'antigravity'.

New exact time-dependent field solutions of Einstein's equation [7] provide further support for the results reported in this paper. The exact field solutions for a mass moving with constant velocity correspond precisely in the weak-field approximation to the fields calculated here for the special case of constant velocity. The complete set of coordinate transformations for relating the exact solutions of [7] to the retarded-field solutions of this paper were presented in Table 1. In Eqs. (17), exact potentials from which the exact solutions of Einstein's equation can be derived in 3-vector notation were also presented here for the first time.

The exact time-dependent gravitational-field solutions of Einstein's equation for a moving mass [7], and the two-step approach [8,9] to calculating exact orbits in dynamic fields, and the retarded fields calculated in this paper all give the same result: Even weak gravitational fields of moving masses are repulsive in the forward and backward directions at source speeds greater than $3^{-1/2}c$.

The weak-field solutions in this paper have potential theoretical and experimental applications in the near term and potential propulsion applications in the long term. In the near term, the solutions can be used to test relativistic gravity in the laboratory. The field solutions, and particularly the 'acceleration field' in Eq. (10), should add to our understanding of how inertia is already embodied in the gravitational field equations of general relativity, and add to our understanding of cosmological models.

In the long term, gravitational repulsion at relativistic speeds opens vistas of opportunities for spacecraft propulsion [8,9]. Particularly appealing is that propulsion of a massive payload to relativistic speeds can be accomplished quickly and with manageable stresses, because the only stresses in acceleration along a geodesic arise from tidal forces much weaker than the propulsion forces.

An interesting challenge is to generalize the approach outlined in [7] to calculate the exact dynamic gravitational fields of *accelerating* masses. The exact result should simplify in the weak-field approximation to the fully relativistic weak field of a mass in arbitrary relativistic motion calculated in Eq. (10). Another interesting challenge is to determine the cosmological implications of the new field solutions presented here and in [7].

I am grateful to Bahram Mashhoon and Sergei Kopeikin for valuable comments and advice.